\documentclass[12pt]{iopart}

\usepackage{graphicx}
\usepackage{epsfig}
\usepackage{bm}
\usepackage{color}
\usepackage{float}
\usepackage{ulem}
\def\be{\begin{equation}}
\def\ee{\end{equation}}
\graphicspath{ {images/} }
\newcommand{\bld}[1]{\boldsymbol{#1}}

\newcommand{\cond}{\mbox{\rule[-0.5em]{.1mm}{1.2em}}}

\begin{document}
\title[Error estimates for the Skyrme-Hartree-Fock model]{Error estimates for the Skyrme-Hartree-Fock model}
\date{Update \today}

\author{ J. Erler$^{1}$ and  P.-G. Reinhard$^{2}$}
\address{$^1$Division Biophysics of Macromolecules, German Cancer Research Center, INF 580, D-69120 Heidelberg, Germany}
\address{$^2$Institut f\"ur Theoretische Physik II, Universit\"at Erlangen-N\"urnberg,
Staudtstrasse 7, D-91058 Erlangen, Germany}

\begin{abstract}
There are many complementing strategies to estimate the extrapolation
errors of a model which was calibrated in least-squares fits.  We
consider the Skyrme-Hartree-Fock model for nuclear structure and
dynamics and exemplify the following five strategies: uncertainties
from statistical analysis, covariances between observables, trends of
residuals, variation of fit data, dedicated variation of model
parameters. This gives useful insight into the impact of the key fit
data as they are: binding energies, charge r.m.s. radii, and charge
formfactor. Amongst others, we check in particular the predictive
value for observables in the stable nucleus $^{208}$Pb, the
super-heavy element $^{266}$Hs, $r$-process nuclei, and neutron stars.
\end{abstract}
\maketitle
%

%\tableofcontents
%\makeatletter
%\let\toc@pre\relax
%\let\toc@post\relax
%\makeatother

\section{Introduction}

This special volume is devoted to error analysis in connection with
nuclear models, particularly those which are calibrated by fits to
empirical data. This paper considers in particular the
Skyrme-Hartree-Fock (SHF) approximation. This is a microscopic model
for nuclear structure and dynamics whose structure can be deduced from
general arguments of low-momentum expansion \cite{(Neg72),Rei94aR}
while the remaining model parameters are determined by adjustment to
empirical data. In early stages, the calibration was more like an
educated search \cite{Bei75a}.  Later developments became increasingly
systematic searches \cite{Bar82a,(Ton84)}. A first straightforward
least-squares ($\chi^2$) fit with estimates of extrapolation errors
was used in \cite{Fri86a}. In the meantime, parametrizations have been
steadily developed further including more and more data and exploiting
the benefits of the $\chi^2$ techniques, for recent examples see,
e.g., \cite{(Klu09),(Kor10)}. The availability of these thoroughly
fitted parametrizations allowed a series of extensive studies of
correlations within the models using covariance analysis which
revealed interesting inter-relations between symmetry energy, neutron
radius, and dipole spectra \cite{(Pie12),(Naz13),(Rei13),(Rei13a)}. In
this paper, we want to discuss error analysis from a more general
perspective. The basic principles have been detailed in
\cite{Dob14a}. We will exemplify a couple of the strategies outlined
there for the case of SHF.  We will chose as test observables partly
standard observables from stable nuclei, e.g. giant resonances in
$^{208}$Pb, and partly far reaching extrapolations to $r$-process
nuclei, super-heavy elements and neutron stars. The combination of
strategies provides interesting insights into the predictive value of
SHF for these observables. A particular and new aspect in this
analysis, not much considered so far, is the variation of input data
which allows to explore the impact of fit data on the parametrizations
and with it on extrapolations.

The paper is outlined as follows:
In section \ref{sec:estimates}, we briefly summarize the
needed formula of statistical analysis and the various strategies for
estimating extrapolation errors.
In section \ref{sec:illust}, we exemplify the chosen strategies step
by step.

\section{Fit of model parameter and error estimates}
\label{sec:estimates}

\subsection{Quality measure and optimization}
\label{sec:quality}

The paper \cite{Dob14a} contains a very detailed explanation of
least-squares ($\chi^2$) fits and related error analysis. We repeat
here briefly the basic formula. It is typical for nuclear
self-consistent mean-field models that one can motivate their formal
structure by microscopic considerations as, e.g., low-momentum
expansion \cite{Rei94aR,(Ben03)}. But the model parameters
$\bld{p}=(p_1...p_{N_p})$ remain undetermined. They are calibrated to
experimental data. To that end, one selects a representative set of
observables $\{\hat{\mathcal{O}}_i,i=1...N_d\}$, typically gross
properties of the nuclear ground state as binding energies and radii.
Then one proceeds along the standard scheme of statistical $\chi^2$
analysis.  We define the quality function
as~\cite{(Bra97a),(Bevington),(Tar05)}
\begin{equation}\label{chi2}
\chi^2(\bld{p})
=
\sum_{i=1}^{N_d}
\frac{\left(\mathcal{O}_i(\bld{p})
      -
      \mathcal{O}^\mathrm{exp}_i\right)^2}
     {\Delta\mathcal{O}_i^2} 
\;,
\end{equation}
where $\mathcal{O}_i(\bld{p})$ stands for the calculated values,
$\mathcal{O}^\mathrm{exp}_i$ for experimental data, and
$\Delta\mathcal{O}_i$ for adopted errors. The best-fit model
parameters $\bld{p}_0$ are those for which $\chi^2$ becomes the
minimum, i.e. $\chi^2_0=\chi^2(\bld{p}_0)=\chi^2_\mathrm{min}$.  The
adopted parameters should be chosen such that $\chi^2_0=N_d-N_p$ which
is the number of degrees of freedom in the fit, see \cite{Dob14a}.

Not only the minimum alone, but also some vicinity
represents a reasonable reproduction of data.  Assuming a statistical
distribution of errors, one can deduce a probability distribution
$W(\bld{p})\propto\exp(-\chi^2(\bld{p}))$ of reasonable model
parameters.  Their domain is characterized by
$\chi^2(\bld{p})\leq\chi^2_0+1$ (see Sec.~9.8 of
Ref.~\cite{(Bra97a)}). Its range is usually small and we can expand
\begin{eqnarray}\label{chi2a}
  \chi^2(\bld{p})
  &\approx&
  \chi^2_\mathrm{0}
  +
  \sum_{\alpha,\beta=1}^{N_p} (p_\alpha-p_{0,\alpha})(\mathcal{C}^{-1})_{\alpha\beta}(p_\beta-p_{0,\beta}),
\\
  (\mathcal{C}^{-1})_{\alpha\beta}
  &=&
  {\textstyle\frac{1}{2}}\partial_{p_\alpha}\partial_{p_\beta}\chi^2\cond_{\,\bld{p}_0}
  {\simeq}
  \sum_i{J}_{i\alpha}{J}_{i\beta}
  \;,
\\
  {J}_{i\alpha}
  &=&
  \frac{\partial_{p_\alpha}\mathcal{O}_i\cond_{\,\bld{p}_0}}{\Delta\mathcal{O}_i}
  \;,
\end{eqnarray}
%\marginpar{{B}}
where $\hat{J}$ is the {rescaled} Jacobian matrix and $\mathcal{C}$ the
covariance matrix. The latter plays the key role in covariance
analysis. The domain of reasonable parameters is thus given by
$\bld{p}\cdot\hat{\mathcal{C}}^{-1}\cdot\bld{p}\leq 1$ which defines a
confidence ellipsoid in the space of model parameters. It is related
to the probability distribution
\cite{(Bra97a),(Tar05)}
\begin{equation}
  W(\bld{p})
  \propto
  \exp(-\frac{1}{2}\bld{p}\cdot\hat{\mathcal{C}}^{-1}\cdot\bld{p})
  \quad.
\label{eq:probab}
\end{equation}

Any observable $A$ is a function of model parameters
$A=A(\bld{p})$. The value of $A$ thus varies within the confidence
ellipsoid, and this results in some uncertainty $\Delta A$. Usually,
one can assume that $A$ varies weakly such that one can linearize it
\begin{equation}
\label{eq:linear}
  A(\bld{p})
  \simeq
  A_0+\bld{G}^A\cdot(\bld{p}- \bld{p}_0)
  \quad\mbox{for}\quad
  A_0
  =
  A(\bld{p}_0)
  \quad\mbox{and}\quad
  \bld{G}^A
  =
  \bm{\partial}_{\bm{p}}A\cond_{\,\bld{p}_0}
  \quad.
\end{equation}
This assumption will be used throughout the paper, with exception of
section \ref{sec:beyondlinear} where we check non-linear effects. 

%Note also that the Jacobians ${J}_{i\alpha}$ are particular slopes
%$G^i_\alpha$ when identifying $A=\hat{\mathcal{O}}_i$.

The covariance matrix $\hat{\mathcal{C}}$ and the slopes $\bld{G}^A$
are the basic constituents of error estimates and correlations within
statistical analysis addressed in the following.

\subsection{Strategies for estimating errors}
\label{sec:estim}

A $\chi^2$ fit is a black box. One plugs in a model, chooses a couple
of relevant fit data, and grinds the mill until one is convinced to
have found the absolute minimum $\chi^2_0$ together with the optimal
parameters $\bld{p}_0$. What remains is to understand the model thus
achieved, in particular its reliability in extrapolations to other
observables. This is the quest for error estimates which does not have
a simple and unique answer. One can only approach the problem from
different perspective and so piece-wise put together an idea of the
various sources of uncertainty.  This has been discussed extensively
from a general perspective in \cite{Dob14a}. We will exemplify that
here on some of the proposed strategies for the particular case of the
Skyrme-Hartree-Fock (SHF) approach.  We assume that SHF is
sufficiently well known to the reader and refer for details to the
reviews \cite{(Ben03),Sto07aR,(Erl11)}.

The strategies for evaluating properties of the model and its
uncertainties are here summarized in brief:
\begin{enumerate}
  \item\label{it:extrap} {\bf Extrapolation uncertainties from
    statistical analysis.}\\ 
    Using the probability distribution (\ref{eq:probab}), one can
    deduce the uncertainty on the predicted value $A_0$ as
   \begin{equation}\label{predicted_error}
     \Delta A
     =
     \sqrt{\overline{\Delta A^2}}
     \;,\;
     \overline{\Delta A^2}
      =
     \sum_{\alpha\beta}G^A_\alpha{\mathcal{C}}_{\alpha\beta}G^A_\beta
     \quad.
   \label{eq:uncert}
   \end{equation}
   This is the statistical extrapolation error serving as useful
   indicator for safe and unsafe regions of the model. It will be
   exemplified in section \ref{sec:uncert}.

  \item\label{it:correl} {\bf Correlations between observables from
    statistical analysis.}\\ 
    Again using $W(\bld{p})$ from Eq.~(\ref{eq:probab}), one can
    deduce also the correlation, or covariance, between two
    observables $A$ and $B$ as
  \begin{equation}
    {c}_{AB}
    =
    \frac{|\overline{\Delta A\,\Delta B}|}
         {{{\Delta A}_0\,{\Delta B}_0}}
   \quad.
   \label{eq:correlator}
   \end{equation}
   A value ${c}_{AB}=1$ means fully correlated where
   knowledge of $A(\bld{p})$ determines  $B(\bld{p})$. 
   A value ${c}_{AB}=0$ means uncorrelated, i.e.
   $A(\bld{p})$ and $B(\bld{p})$ are statistically independent. 
   We will exemplify covariance analysis in section \ref{sec:covar}.

  \item\label{it:sensit} {\bf Sensitivity analysis for the model
    parameters $\bld{p}$.}\\ The Jacobian matrix $\hat{J}$ together
    with the covariance matrix $\hat{\mathcal{C}}$ allows to explore
    the impact of each single model parameter $p_\alpha$ on a given
    fit observable $\hat{\mathcal{O}}_i$. Examples for this kind of
    analysis are found in \cite{(Kor10),(Kor12),Dob14a}.

  \item\label{it:trends} {\bf Dedicated variations of parameters.}\\
    It can be instructive to watch the trend of an observable
    $A(\bld{p})$ when varying one parameter $p_\alpha$. This becomes
    particularly useful when expressing the SHF model parameters
    in terms of bulk properties of nuclear matter.
    This strategy will be exemplified in section \ref{sec:varyNMP}.  

  \item\label{it:reserr} {\bf Trends of residual errors.}\\ 
    A perfect model should produce a purely statistical (Gaussian)
    distribution of residuals
    $\mathcal{O}_i(\bld{p})-\mathcal{O}^\mathrm{exp}_i$.  Unresolved
    trends indicate deficiencies of the model.  All nuclear mean-field
    models produce still strong unresolved trends, see, e.g.,
    \cite{(Klu09),(Kor10),(Erl11)}. We will discuss a compact
    version of this analysis in section \ref{sec:repro}.
     
  \item\label{it:varfit} {\bf Variations of fit data.}  The fit
    observables $\hat{\mathcal{O}}_i$ stem from different groups of
    observables as, e.g., energy or radius.
    One can omit this or that group from the pool of fit
    data. Comparison with the full fit allows to explore the impact of
    the omitted group. This strategy will be initiated in section
    \ref{sec:varydata} and carried forth throughout the paper in
    combination with the other strategies.

  \item\label{it:compar} {\bf Comparison with predicted data. }\\ 
    A natural test is to compare a prediction
    (extrapolation) with experimental data. The model is probably
    sufficient if the deviation remains within the extrapolation error
    (\ref{eq:uncert}). It is a strong indicator for a systematic error
    if this is not the case. An example is the energy of
    super-heavy nuclei where deviations (and residuals) point to
    a systematic problem of the SHF model \cite{Erl10a}.

  \item\label{it:varmod} {\bf Variations of the model.}
    The hardest part in modeling is to estimate the systematic
    error. All above strategies explore a model from within. This
    gives at best some indications for a systematic error. For more,
    one has to step beyond the given model.  One way is to
    extend the model by further terms. 
    Another way is to compare and/or accumulate the
    results from different models. Examples for the latter strategy
    can be found in \cite{(Pie12),(Rei13a)}.

\end{enumerate}
{ As discussed in \cite{Dob14a}, there is a basic distinction
  between statistical error and systematic error. Quantities related
%\marginpar{{Ref 2}}
  to the statistical error can be evaluated within the given model,
  here SHF, and quality measure $\chi^2$. This concerns points
  \ref{it:extrap}--\ref{it:trends} in the above list, to some extend
  also point \ref{it:varfit}. It is a valuable tool to estimate a
  lower limit for extrapolation errors. The systematic error, on the
  other hand, covers all insufficienties of the given model. There is
  no systematic way to estimate as we usually do not dispose of the
  exact solution to compare with. It can only be explored piece-wise
  from different perspectives. These are the methods in points
  \ref{it:varfit}--\ref{it:varmod}. Each one of the methods
  \ref{it:reserr}--\ref{it:varmod} amounts to a large survey of the
  validity of SHF on its own. Thus we refer in this respect to the
  papers already mentioned in the list above.}

\section{Results and discussion}
\label{sec:illust}

\subsection{{Choices for the variation} of input data}
\label{sec:varydata}

First, we take up point \ref{it:varfit} of section \ref{sec:estim},
the variation of fit data. {We explain in this section the
%\marginpar{{C}}
  strategy for variations and choice of input data. The
  parametrization thus obtained under the varied conditions will then
  be used throughout all the following sections. The variation of
  input data unfolds only in connection with the subsequent strategies.
}

Basis is the pool of fit data as developed
and used in \cite{(Klu09)}. It contains binding energies, form
parameters from the charge formfactor (r.m.s. radius $r_\mathrm{rms,C}$,
diffraction radius $R_\mathrm{diffr,C}$, surface thickness $\sigma_C$
\cite{Fri82a}), and pairing gaps $\Delta_\mathrm{pair}$ in a selection
of semi-magic nuclei which had been checked to be well reproduced by
mean-field models \cite{(Klu08)}. Furthermore, it includes some
spin-orbit splittings $\varepsilon_\mathrm{ls}$ in doubly magic
nuclei. An unconstrained fit to the full set yields the
parametrization SV-min. Now we have performed a series of fits with
deliberate omission of groups of fit data.
%\begin{SCtable}[5]
\begin{table}[bht]
\begin{center}
\begin{tabular}{|l|c|ccc|cc|}
\hline
  & $E_B$  &  $r_{\mathrm{rms,C}}$ & $R_{\mathrm{diff,C}}$ & $\sigma_\mathrm{C}$ &
 $\varepsilon_\mathrm{ls}$  &  $\Delta_\mathrm{pair}$ \\
\hline
 SV-min &   x & x & x & x &  x &  x \\
 ``no surf'' &   x & x & x & - &  x &  x \\
 ``no rms'' &  x & - & x & x &  x &  x \\
 ``no Formf'' &  x & x & - & - &  x &  x \\
 ``E only''  &   x & - & - & - &  x &  x \\
\hline
\end{tabular}
\end{center}
\caption{\label{tab:names} Included data sets from the standard pool
  of fit data from finite nuclei \cite{(Klu09)}.  A ``x'' means
  included and ``-'' stands for excluded.  The observables are
  explained in the text.  }
\end{table}
%\end{SCtable}
%
This yields the parametrizations as listed in table \ref{tab:names}.
We have also studied omission of $\varepsilon_\mathrm{ls}$ or
$\Delta_\mathrm{pair}$. This showed only minor effects and so we do
not report on these variants.
{The effects of the various omissions on the average deviations
  in the first four blocks of data will be shown later
  in figure \ref{fig:collect-errorblocks} and discussed at that place.}

\subsubsection{Effect on predicted observables}

\begin{figure}
\centerline{\epsfig{figure=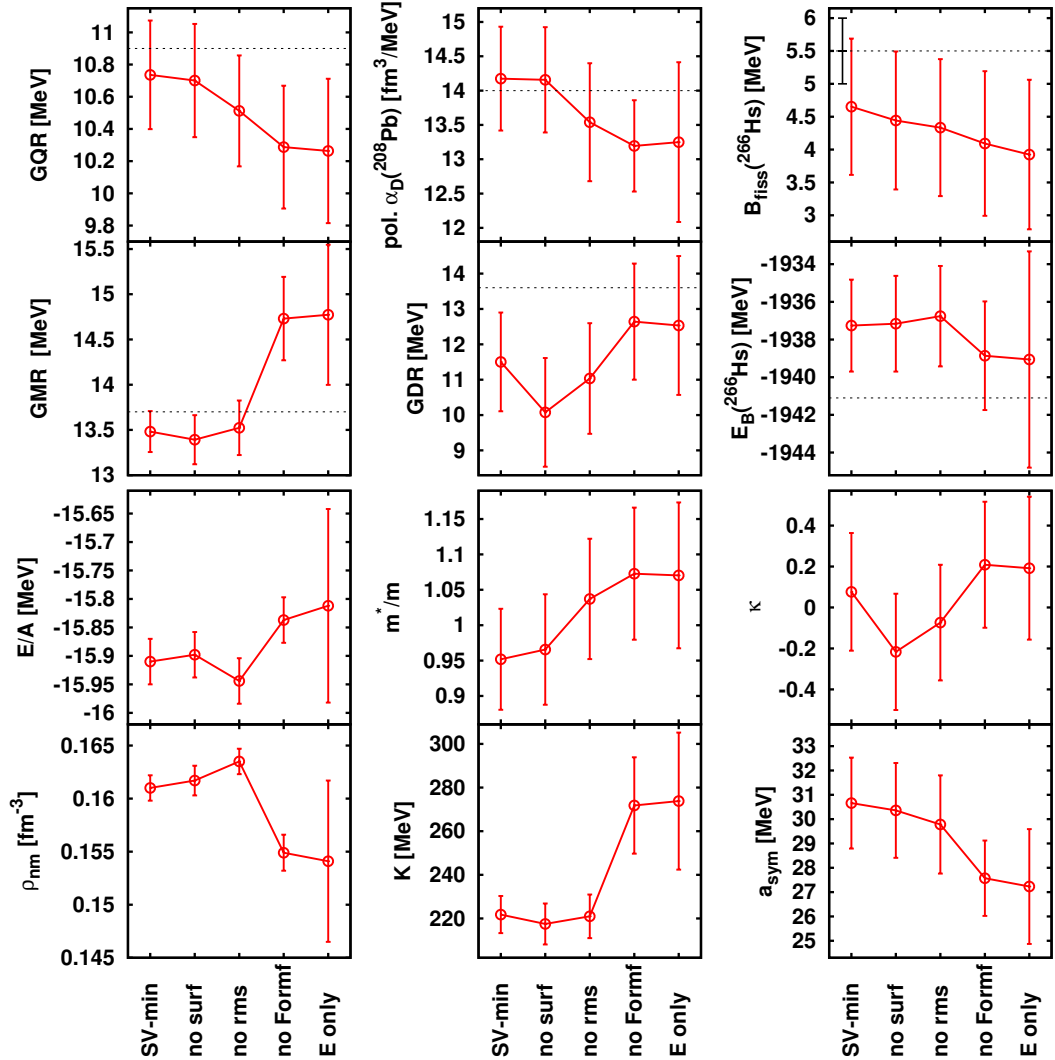,width=0.9\linewidth}}
\caption{\label{fig:collect-NMP} Lower 6 panels: Predicted
  nuclear-matter properties (NMP) with their extrapolation
  uncertainties for the parametrizations in table \ref{tab:names}.
  Upper 6 panels: Predicted properties in finite nuclei, giant
  resonance energies and dipole polarizability $\alpha_D$ in
  $^{208}$Pb and binding energy $E_B$ as well as fission barrier $B_f$
  in $^{266}$Hs.  The faint dotted horizontal lines indicate the
  experimental data.  }
\end{figure}
In this section, we look at the effect of varied fit data on
predicted/extrapolated observables, nuclear matter properties (NMP)
and key observables of the two nuclei $^{208}$Pb and $^{266}$Hs. In
$^{266}$Hs, we consider the binding energy $E_b$ \cite{Wan12} and the
fission barrier $B_f$ \cite{Pet04}.  The experimental value for $B_f$
is augmented with an error bar as there remains a large uncertainty
from experimental analysis.  The binding energy include the rotational
zero-point energy which is obligatory for deformed nuclei
\cite{(Klu08)}. The fission barriers are taken relative to the
collective ground state and include rotational correction
\cite{Sch09a,Erl12a}. {Note that we thus include for this
  observable some correlation effects beyond SHF which means that
%\marginpar{{Ref 2}}
  $B_f$ is not a pure mean-field observable.} In $^{208}$Pb, we
consider the isoscalar giant monopole (GMR) and quadrupole (GQR)
resonance, the isovector giant dipole resonance (GDR), and the dipole
polarizability $\alpha_D$. These four observables are computed with
the techniques of \cite{Rei92}.
The NMP considered are: binding energy $E/A$, density $\rho_{nm}$,
incompressibility $K$, isoscalar effective mass $m^*/m$, symmetry
energy $a_\mathrm{sym}$, and Thomas-Reiche-Kuhn (TRK) sum rule
enhancement factor $\kappa_\mathrm{TRK}$, all taken at the equilibrium
point of symmetric matter. Note that NMP can be viewed as observables
$\hat{A}$ and equally well as model parameters $\hat{A}\equiv
p_\alpha$. This is no principle difference. All formula for
uncertainty and correlations can be employed when identifying
$G^{p_\beta}_\alpha=\delta_{\alpha\beta}$. Figure~\ref{fig:collect-NMP}
shows the effect of the omissions of data on the observables and their
uncertainties.  Changes in uncertainty indicate the impact of the
omitted data group an the observable.  A shift of the average shows
what data are pulling in which direction.

The effects on NMP (lower six panels) are large throughout.  A most
pronounced shift is produced by omitting $R_\mathrm{diffr,C}$ (in ``no
Formf'' and ``E only'') which leads to a large jump in bulk
equilibrium density $\rho_\mathrm{nm}$ and incompressibility
$K$. There is also a jump in the isovector response
$a_\mathrm{sym}$ in addition to the generally strong changes.  It is
also interesting to note that radius information keeps the effective
mass $m^*/m$ down to values below 1 while fits without radii let
$m^*/m$ grow visibly above one.
{The reason is probably that $m^*/m$ has an impact on the surface
%\marginpar{{E}} 
profile, thus on $r_\mathrm{rms,C}$ and
$\sigma_\mathrm{C}$, and, in turn, also on $R_\mathrm{diffr,C}$. }
The variances, of course, grow generally when omitting data.  A
particularly large increase emerges for the set ``E only''.  This
indicates that any information from the charge formfactor is extremely
helpful to confine the parametrization. 
On the other hand, {although the error
%\marginpar{{A}}  
bars for ``E only'' are generally larger than
for the other parametrizations, they are still in acceptable
ranges. This shows that energy data} alone can already provide a
reasonable parametrization.

The upper block in figure \ref{fig:collect-NMP} shows the effect of
omission of fit data on observables in finite nuclei. The three giant
resonances and the polarizability $\alpha_D$ are known to have a
one-to-one correspondence with each one NMP \cite{(Klu09)}: the GMR
with $K$, the GQR with $m^*/m$, the GDR with $\kappa_\mathrm{TRK}$,
and $\alpha_D$ with $a_\mathrm{sym}$. These pairs of highly correlated
observables shows the same trends. 
%\marginpar{{F}}  
{Note that this one-to-one correspondence is
  obtained in the dedicated variation of one NMP when re-optimizing
  all other NMP (as explained in section \ref{sec:varyNMP}); isolated
  variation can show more dependences as, e.g., a dependence of the
  GDR on $a_\mathrm{sym}$ \cite{(Tri08)}.  Note, furthermore, that
  the correspondence of $\alpha_D$ with $a_\mathrm{sym}$ is equivalent
  to a correspondence of $\alpha_D$ with $\partial_\rho
  a_\mathrm{sym}$, because $a_\mathrm{sym}$ is strongly correlated
  with $\partial_\rho a_\mathrm{sym}$ \cite{Roc13}.  } The large
effects from omission of $R_\mathrm{diffr,C}$ in ``no Formf'' which
were seen in some NMP come up again here. Unlike NMP, observables in
finite nuclei allow a comparison with experimental data. The giant
resonances (with
%\marginpar{{G}}
exception of GDR) and $\alpha_D$ indicate that inclusion of radius
information drives into the right direction {which is a gratifying
feature.}
Inclusion of
radii is also beneficial for $B_f(^{266}$Hs$)$ but disadvantageous for
$E_B(^{266}$Hs$)$. It is noteworthy that the trend of
$E_B(^{266}$Hs$)$ is similar to the trend of $\rho_\mathrm{nm}$.  This
indicates some correlation between these two observables which is seen
in figure \ref{fig:alignmatrix-both}.

Besides the strong impact of $R_\mathrm{diffr,C}$, there remain only
few minor effects. The set ``E only'' produces again a large growth of
the variance for the energy observable $E_B(^{266}\mathrm{Hs})$ which
correlates well with a similar huge growth for the bulk $E/A$.

\subsubsection{Reproduction of fit observables}
\label{sec:repro}

In this subsection, we look at the effect of varied fit data on the
average and r.m.s. residuals taken over a group of data. This is a
compact, although already informative, version of the study of
residuals mentioned under point \ref{it:reserr} in section
\ref{sec:estim}. Mean values deviating significantly from zero within
the scale set by the uncertainties indicate a non-statistical
distribution and thus some incompatibility of the observable with the
model. Changes on the r.m.s. error indicate the sensitivity to a group
of observables.
\begin{figure}
\centerline{\epsfig{figure=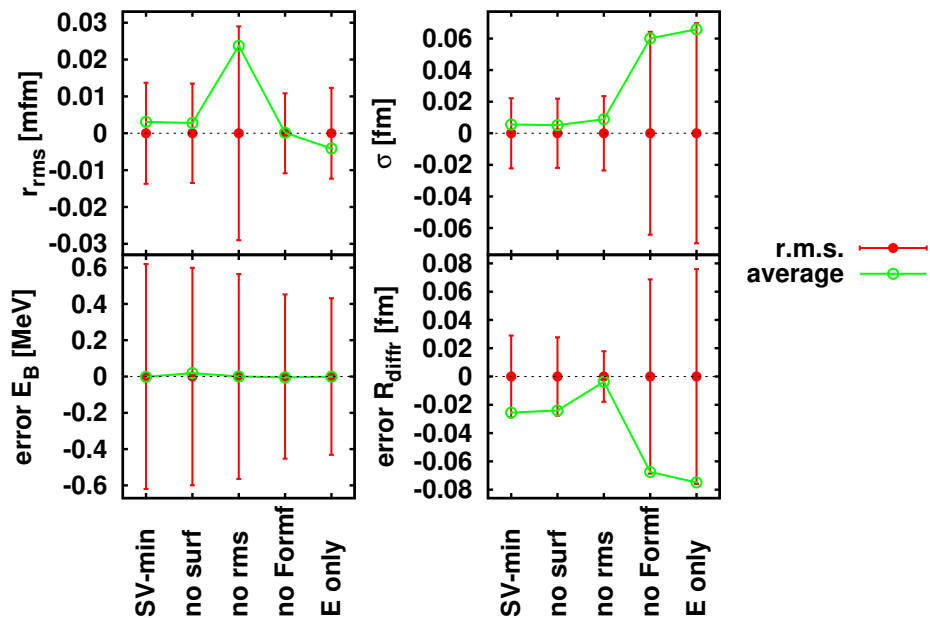,width=0.79\linewidth}}
\caption{\label{fig:collect-errorblocks}
Average deviation (green) and r.m.s. deviation (red errorbars) for
groups of observables in the pool of fit data.
}
\end{figure}
Figure \ref{fig:collect-errorblocks} shows the effects of the
omissions on the average residuals of fit observables. 
The average error of energy $E_B$ is always near zero which is not
surprising because all fits include $E_B$ with large weight. Note that
the zero average error does not exclude missing trends which are seen
when plotting the systematics of energy deviations, see
e.g. \cite{Erl10a,(Klu09),(Kor10)}.

The diffraction radius $R_\mathrm{diffr,C}$ is special in that it
shows large average residuals for practically every case. In fact, the
r.m.s.  error is almost exhausted by the average error. Zero average
error appears only for the set ``no rms'' which omits
$r_\mathrm{rms,C}$. This indicates that there is some incompatibility
between $r_\mathrm{rms,C}$ and $R_\mathrm{diffr,C}$ in the present
model. Negative average error means that the $R_\mathrm{diffr,C}$ tend
to be larger than the experimental values. The opposite trend is seen
for $r_\mathrm{rms,C}$ (upper left panel, case ``no rms'') showing
that $r_\mathrm{rms,C}$ wants to be smaller than the data if
$R_\mathrm{diffr,C}$ is matching perfectly. The sets ``no Formf'' and
``E only'' show where $R_\mathrm{diffr,C}$ ends up if it is not
constrained by fit. 
%The average deviation is far outside the r.m.s. error of SV-min.
{The change of its average deviation (green lines) 
%\marginpar{{A}}
(and with it the average values) is significant. It grows to values
of about $0.07\,\mathrm{fm}^{-3}$ which is far larger than the
allowed uncertainty from SV-min of about  $0.02\,\mathrm{fm}^{-3}$
(red error bars). Thus} 
 the reproduction of the charge formfactor is
much degraded for these two sets. The question is whether this is an
insufficiency of the SHF model or whether we see here a defect of the
model for the intrinsic nucleon formfactor \cite{Fri86a}. 
Note that the deviations
%\marginpar{{A}} 
are of the order of the adopted errors {(0.04 fm$^{-3}$).  This
  indicates that a better simultaneous adjustment of $E_B$ and
  $R_\mathrm{diffr,C}$ is not possible within the given SHF model.}
The trend {of the average deviations of} $R_\mathrm{diffr,C}$ is
very similar to the trend of $\rho_\mathrm{nm}$ in figure
\ref{fig:collect-NMP}.  The correlation goes up to quantitative
detail:  From SV-min to ``E only'', $R_\mathrm{diffr,C}$ grows by 0.05
fm, i.e. by about 1\%. The density $\rho_\mathrm{nm}$ shrinks by about
0.005 fm$^{-3}$ which corresponds perfectly to the radius effect.

The surface thickness $\sigma_\mathrm{C}$ has generally a small
average error. Omitting only $\sigma_\mathrm{C}$ does not change much.
The correct $\sigma_\mathrm{C}$ emerges already if
$R_\mathrm{diffr,C}$ is properly constrained.  We see, however, a
drive to smaller surfaces if all formfactor observables are omitted in
the fit.

\subsection{Extrapolation uncertainties}
\label{sec:uncert}

According to point \ref{it:extrap} in section \ref{sec:estim}, the
most obvious result of statistical analysis are the uncertainties on
extrapolated observables. It is natural that extrapolations become the
more risky the farther away a nucleus is from the fit pool. 
\begin{figure}[htb]
\center
\includegraphics[width=0.79\textwidth]{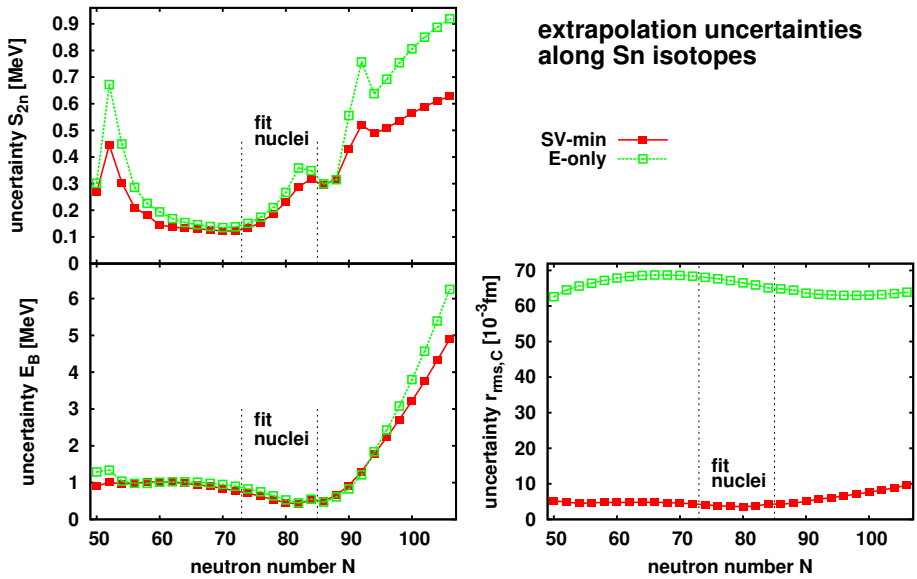}
\caption{\label{extrap-chain2} Extrapolation uncertainties for binding
  energy $E_B$ (lower left panel), charge r.m.s. radius $r_\mathrm{rms,C}$
  (lower right
  panel), and two-neutron separation energy $S_{2n}$ (upper left
  panel) along the chain of even Sn isotopes. The few Sn nuclei which
  were included in the fits lie in between the two vertical dotted
  lines. Results are shown for two forces, SV-min and ``E only''.
}
\end{figure}
This is demonstrated in figure \ref{extrap-chain2} for the chain of Sn
isotopes reaching out to very neutron rich $r$-process nuclei. The
uncertainties of the fit observables $E_B$ and $r_{rms,C}$ are
smallest for the fit nuclei and grow with distance to the fit
region. The growth is large, by a factor of 6--10, for the
extrapolation of $E_B$ and of two-neutron separation energy $S_{2n}$
into the neutron-rich region. This is a direction where
isovector terms become important, but isovector NMP
are not so well fixed in fits to nuclear ground states.
Extrapolations to superheavy elements are probing more the isoscalar
channel and are thus more robust showing only factor 2--3 growth in
uncertainty, see $E_B(^{266}\mathrm{Hs})$ in figure
\ref{fig:collect-NMP}. The fit ``E only'' has comparable uncertainties
on $E_B$ and $S_{2n}$ for the fit nuclei, but shows slightly faster
growth outside this region. The effect is not dramatic which indicates
that fits to energy suffice to predict energy observables.  Different
is the observable $r_{rms,C}$. The fit ``E only'' has an order of
magnitude larger uncertainties than SV-min. However, both forces agree
in the trend which is very flat over the whole chain. Predictions for
radii thus are robust once radii are well fitted.

\begin{figure}
\centerline{\epsfig{figure=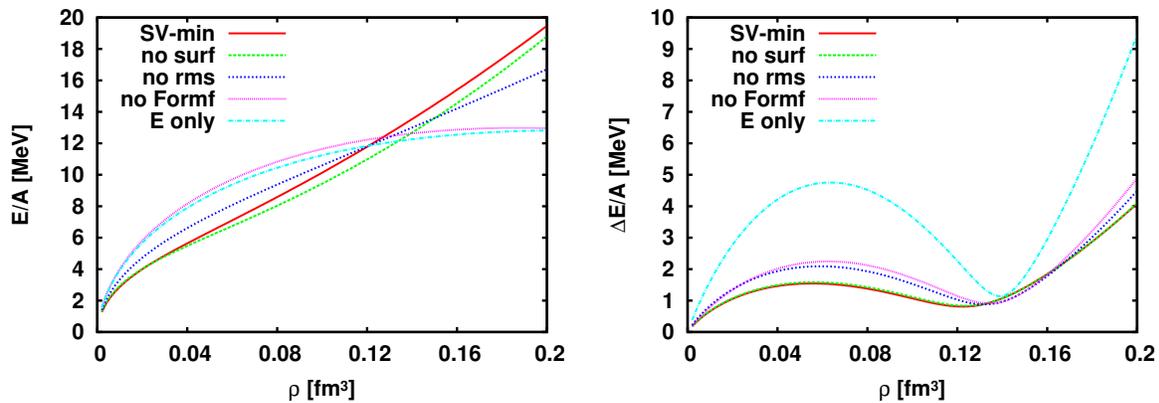,width=0.99\linewidth}}
\caption{\label{fig:EoSneut-SVmin-variations} Left: $E/A$ versus
  density for pure neutron matter for the parametrization developed
  with the fit protocol of SV-min while omitting certain blocks of
  observables.  Right: The extrapolation uncertainties for the data
  shown in the left panel.  }
\end{figure}
A very far extrapolation is involved in studying pure neutron matter
as it is often done in nuclear astrophysics (see
\cite{LattRev,Sto07aR,(Erl13)} and citations therein). The left panel
of figure \ref{fig:EoSneut-SVmin-variations} shows the equation of
state (EoS) $E/A_\mathrm{neut}(\rho)$ for pure neutron matter. There
arise huge differences in its slope {at $\rho<0.12\,\mathrm{fm}^{-3}$
and  $\rho>0.15\,\mathrm{fm}^{-3}$.}   In spite of these very
different slopes, all EoS share about the same values 
%\marginpar{{A}}
{around neutron density $\rho\approx{0.13}$ fm$^{-3}$.
It looks like a ``fixpoint'' in the neutron EoS.}

The right panel shows the corresponding extrapolation
uncertainties. The errors follow the same trends as the deviations
between the parametrizations in the left panel. In regions of large
deviations, these  are larger than the estimated uncertainties.  The
discrepancy indicates that systematic errors will play a role here.
What the $\rho$-dependence of the uncertainties is
concerned, it is astonishing that there arises a pronounced minimum
just near this ``magic'' density $\rho\approx{0.13}$ fm$^{-3}$ where
all predictions approximately agree. Although, the actual position of
the minimum varies a bit with the force the coincidence looks
impressive.  The reasons for this particularly robust point has yet to
be found out.

The {difference in neutron EoS, and particularly the difference
in slope for  $\rho<0.12\,\mathrm{fm}^{-3}$,}
has dramatic consequences for the stability
of neutron stars. The two forces with small slope ``no Formf'' and ``E
only'' do not yield a maximal radius at all because the neutron EoS
becomes unbound for very large densities.

%\clearpage

\subsection{Variation of nuclear matter properties (NMP)}
\label{sec:varyNMP}

We now come to point \ref{it:trends} of section \ref{sec:estim}, the
dedicated variation of model parameters. We do that in terms of the
NMP which are a form of model parameters with an intuitive physical
content. We consider here a variation of only one NMP at a time (this
differs from the variation in \cite{(Klu09)} where four NMP were kept
fixed). Thus we keep only the one varied NMP at a dedicated value and
fit all remaining, now 13, model parameters. This is so to say a
correlated variation because it allows the other model parameters to
find their new optimum value for the one given constrained NMP.

\begin{figure}
\center
\includegraphics[width=0.8\textwidth]{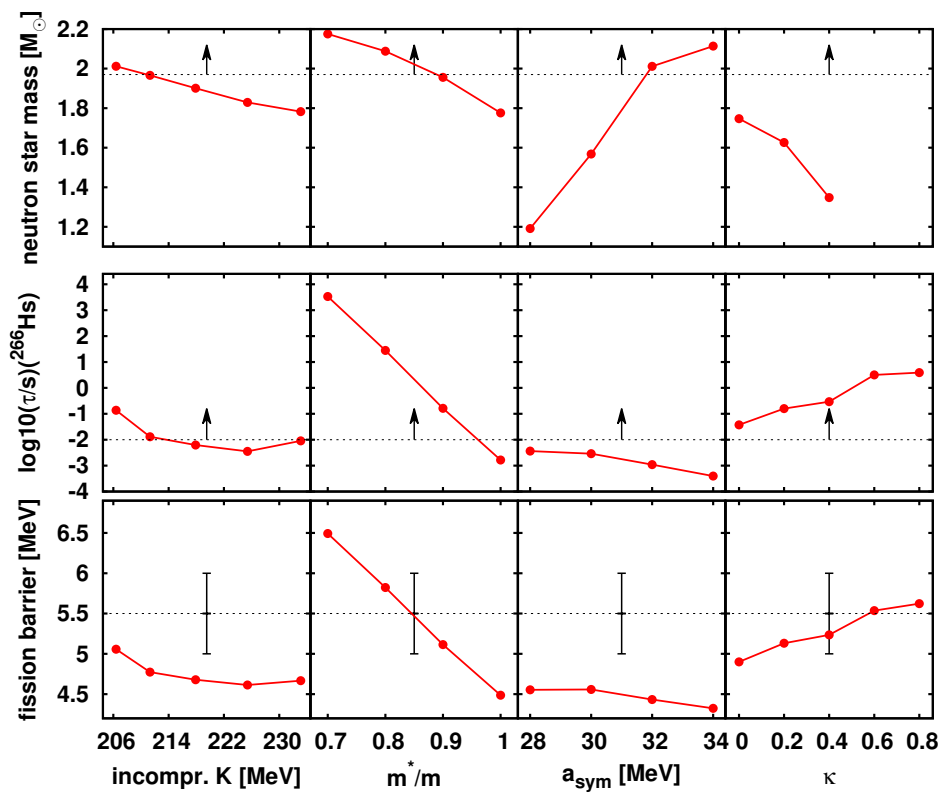}
\caption{\label{collect-neustar-fiss} Results from systematic 
variation of NMP. The maximal mass of neutron stars (upper), fission 
lifetime (middle) and fission barrier (lower) in $^{266}$Hs as 
function of the four varied NMP ($K$, $m^*/m$, 
$a_\mathrm{sym}$, or $\kappa_\mathrm{TRK}$). One NMP was fixed while performing 
free fit of all other parameters. In cases where data is 
missing for the neutron star mass, the equation of state is not 
stable for high densities (see figure~\ref{fig:EoSneut-SVmin-variations}).
Dashed lines show experimental data for the maximal neutron star 
mass \cite{Dem10a}, the lifetime \cite{Hof01} and the barrier \cite{Pet04} 
of $^{266}$Hs. Up-arrows indicate that these data are just lower limits.}
\end{figure}
Figure~\ref{collect-neustar-fiss} shows the maximal neutron star mass
obtained by solving the Tolman-Oppenheimer-Volkoff equation (see
\cite{(Erl13)} for details) as well as fission barriers and fission
lifetimes of the super-heavy nucleus $^{266}$Hs. Fission barriers were
determined with respect to the collective ground state energy. The
collective ground state and the lifetime were calculated using the
procedure as described in \cite{Erl12a}. The incompressibility $K$ has
little effect on all three observables while the effective mass
$m^*/m$ shows always strong trends. The effect on $^{266}$Hs is
understandable because $m^*/m$ determines the spectral density and
with it the shell-corrections which are known to have a strong
influence on the fission path. Different behaviors are seen for
$a_\mathrm{sym}$ and $\kappa_\mathrm{TRK}$. Neutron matter depends
sensitively on these isovector parameters while fission in $^{266}$Hs
reacts less dramatic, although there remains a non-negligible trend
also here. However strong or weak the trends, all three observables
gather influences from several NMP. Unlike the case of giant resonances
\cite{(Klu09)}, there is here no one-to-one correspondence between an
observable and one NMP.

\subsection{Covariances}
\label{sec:covar}

Statistical correlations, also called covariances ${c}_{AB}$ between
two observables $A$ and $B$, see point \ref{it:correl} of section
\ref{sec:estim}, are a powerful tool to explore the hidden connections
within a model.
%A high correlation can be read in two ways. First, it says that the
%measurement of one observable practically determines the value of the
%other. Second, the same strong prediction can be taken as a test of a
%model. There is some inconsistency if we have a strong correlation but
%find that one of the predicted values is far from the data.
This quantity helps, e.g., to determine the information content of a
new observable added to an existing pool of measurements.  Examples
and detailed discussions are found in previous papers
\cite{Rei10,(Rei13),Dob14a}. We add here two new cases related to the
observables and parametrizations addressed in this paper.
\begin{figure}
\centerline{\epsfig{figure=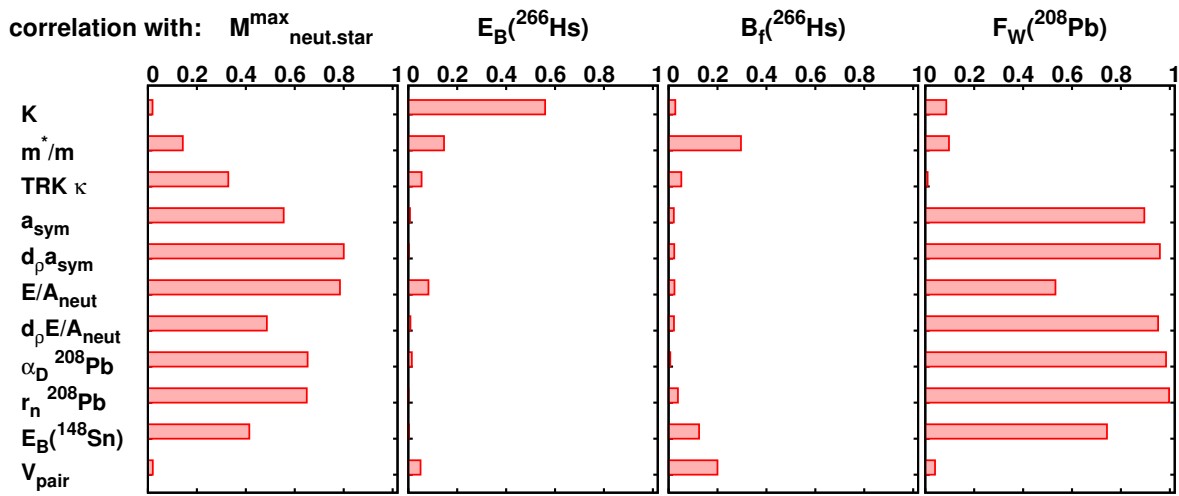,angle=-90,width=\linewidth}}
\caption{\label{fig:aligns} Correlations of the maximal mass
  of a neutron star, of the binding energy and of the fission barrier
  of the isotope $^{266}$Hs, and of the weak-charge formfactor 
  $F_W(q\!=\!0.475/\mathrm{fm})$
  in $^{208}$Pb with NMP and a couple of observables in finite nuclei,
  for the force SV-min. The $E/A_\mathrm{neut}$ stands for the neutron
  EoS at $\rho=0.1$ fm$^{-3}$ and $d_\rho E/A_\mathrm{neut}$ for its
  slope at this density. $\alpha_D$ is the dipole polarizability, and
  $r_\mathrm{n}$ the neutron r.m.s. radius. For the pairing strength, we show
  only the proton case. The correlations for neutron pairing strength
  are equivalent.  }
\end{figure}
Figure \ref{fig:aligns} shows the covariances for four observables,
three of them being far extrapolations and one rather at the safe
side.  The latter case is the weak-charge formfactor $F_W$ in
$^{208}$Pb (right panel), 
%\marginpar{{J}}
{taken at momentum $q\!=\!0.475/\mathrm{fm}$,} which is known to be closely related to the
neutron radius \cite{(Rei13a)}. It is fully correlated with all the
static isovector observables $a_\mathrm{sym}$, $d_\rho
E/A_\mathrm{neut}$, and $\alpha_D$, but uncorrelated with the
isoscalar NMP and the dynamic isovector response
$\kappa_\mathrm{TRK}$.  {As an example for an exotic nucleus
deep in the astro-physical $r$-process region,
%\marginpar{{I}}
we have included $^{148}$Sn in the considerations.  }
The strong isovector correlations provide also
a sizable correlation with this extremely neutron rich $^{148}$Sn. 

The maximal mass of a neutron star (left panel) shows a somewhat more
mixed picture. It is, not surprisingly, most strongly correlated with
the neutron EoS. There are also sizable correlations with all the
static isovector observables {(block from $a_\mathrm{sym}$ to $r_n$)} and to some extend with the extrapolation
to the extremely neutron rich $^{148}$Sn as well as
$\kappa_\mathrm{TRK}$. Very little correlation exists with the
isoscalar NMP $K$ and $m^*/m$.

A much different picture emerges for the binding energy of the
super-heavy $^{266}$Hs: 
%\marginpar{{A}} 
{There is no really large correlation with any
  observable shown here. Negligible are correlations with static
  isovector observables.  Some correlations exit for the parameters
  $K$, $m^*/m$, $\kappa_\mathrm{TRK}$, and $V_\mathrm{pair}$ which
  indicates that all these four parameters have some impact on
  $E_B(^{266}$Hs$)$.}  The fission barrier $B_f(^{266}\mathrm{Hs})$
shows similarly a collection of many small correlations. The most
pronounced here is the correlation with the effective mass
$m^*/m$. This agrees with results in figure
\ref{collect-neustar-fiss}, where only $m^*/m$ shows a significant
trend {for $B_f$}. Next to $m^*/m$ comes the influence from
pairing strength which is not surprising as pairing has a large effect
near the barrier where the density of states is high. 
%\marginpar{{K}}
%foot{Critizised sentence erased.}
%Pairing has some
%small correlation with $E_B(^{266}\mathrm{Hs})$, but is
%negligible for {the formfactor of $^{208}$Pb and for} neutron matter.

We have seen in figure \ref{fig:EoSneut-SVmin-variations} that the
neutron EoS has a pronounced variation of extrapolation
uncertainties as function of $\rho$. In particular, there is a
marked minimum at $\rho\approx$ 0.13 fm$^{-3}$.
\begin{figure}
\centerline{\epsfig{figure=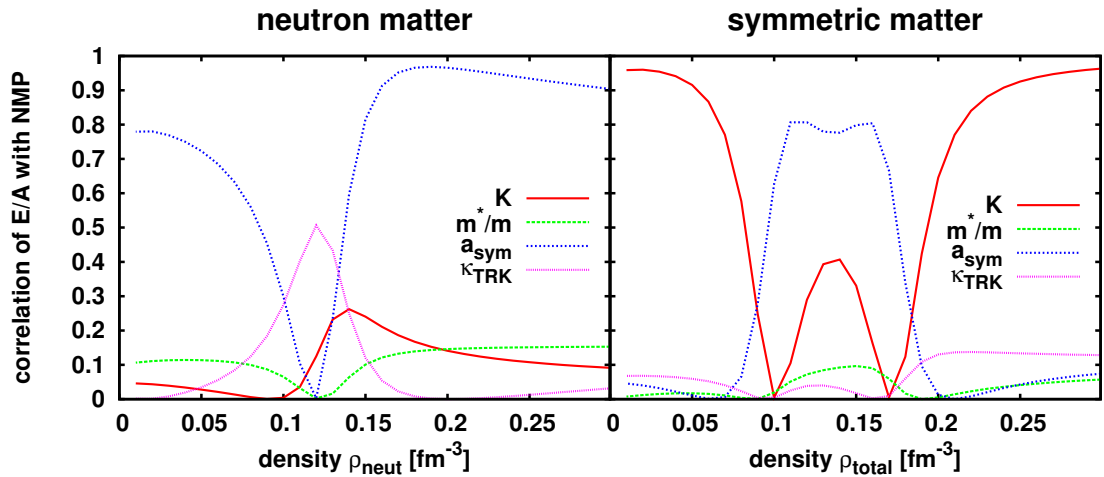,width=0.95\linewidth}}
\caption{\label{fig:E-correl} Correlations of $E/A$ at various 
densities with the basic nuclear matter parameters (NMP). Left: for 
the neutron $E/A_\mathrm{neut}$. Right: for $E/A_\mathrm{sym}$ 
of symmetric nuclear matter. }
\end{figure}
%We take a fresh look at this ``magic'' point by showing 
{As complementing information, we show in the left panel of} figure
\ref{fig:E-correl} the correlation of $E/A_\mathrm{neut}(\rho)$ for
neutron matter with the basic NMP. {Not surprisingly, the symmetry
  energy} $a_\mathrm{sym}$ dominates, at least in the regions of low
and high density. We see a minimum of correlations for all NMP, except
$\kappa_\mathrm{TRK}$, just in the region where uncertainties are
lowest {and where all predictions for $E/A_\mathrm{neut}$ agree,
  see} figure \ref{fig:EoSneut-SVmin-variations}.
%\marginpar{{L}} 
{This is a remarkable coincidence for which we
  have not yet an explanation. }
%This may be
%explain why this region is best predicted. However the maximum of
%$\kappa_\mathrm{TRK}$
%is a bit puzzling and needs yet to be understood.

{The right panel of figure
\ref{fig:E-correl} shows the correlations for $E/A$ of} 
symmetric matter (right panel), $K$ is strongly correlated and
dominates at low and high densities.  
{It is interesting to see that}
%\marginpar{{M}}
$a_\mathrm{sym}$ dominates with sizable correlations in the region
$\rho=$0.10--0.16 fm$^{-3}$.
{This is plausible because these are the typical density values
in the inner surface of a nucleus and this is the region
where the dipole response is predominantly explored.}
The dynamic NMP, $m^*/m$ and $\kappa_\mathrm{TRK}$, are almost
uncorrelated everywhere.

\begin{figure}
\centerline{
\epsfig{figure=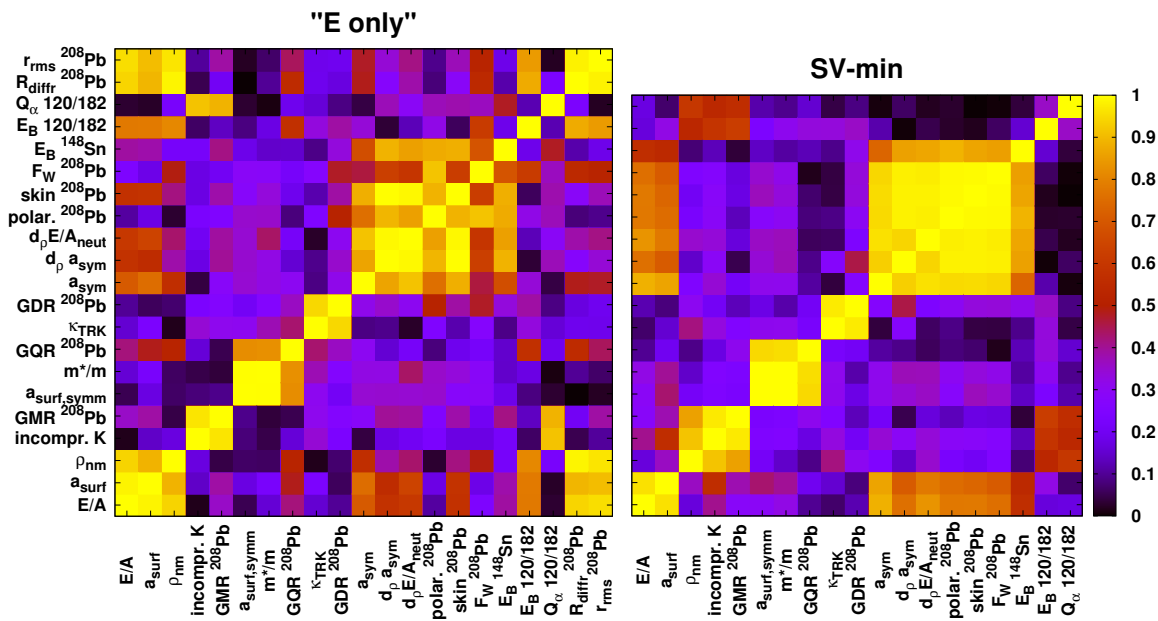,width=\linewidth}
}
\caption{\label{fig:alignmatrix-both}
Correlation matrix for a couple of observables computed from the fit
to all data SV-min (right panel) and  ``E only'' where all information
from radii and surface thickness is omitted (left panel).
High correlation is indicated by light yellow, no correlation by deep
black. 
%Grey dashed boxes indicate observables present in both panels.
}
\end{figure}
Figure \ref{fig:alignmatrix-both} shows the matrix of pairwise
covariances for a selection of NMP and observables in finite nuclei.
A matrix for SV-min (right panel) was already discussed in
\cite{(Kor12)}.  It segregates nicely into four groups of observables:
static isoscalar around $K$, dynamic isoscalar around $m^*/m$, static
isovector around $a_\mathrm{sym}$, and dynamic isovector around
$\kappa_\mathrm{TRK}$. The basic bulk NMP $E/A$ and $\rho_\mathrm{nm}$
are weakly correlated with the static isovector block. The same block
is also weakly correlated with the extremely neutron rich
extrapolation $^{148}$Sn. The far superheavy nucleus Z=120/N=182 has a
bit of low correlations to all other observables, similar to
$^{266}$Hs in figure \ref{fig:aligns}.

The left panel of figure \ref{fig:alignmatrix-both} shows the result
for the set ``E only''. The four blocks of mutually correlated
observables remain almost the same, however, sometimes with somewhat
reduced correlation. A marked change appears for $\rho_\mathrm{nm}$
which was strongly linked to the static isoscalar $K$ block for SV-min
and now moves totally to the basic NMP $E/A$. This case demonstrates
that changing the fit data can have an impact on the
%\marginpar{{N}}
covariances, {and in general does. It is rather surprising that
the two fits show so widely similar trends in covariances.}

The set ``E only'' does not include any radius information in the
data.  This allows to compute also the covariances with charge radii.
The left panel thus includes also $R_\mathrm{diffr,C}$ and
$r_\mathrm{rms,C}$ amongst the observables. Both, $R_\mathrm{diffr,C}$
and $r_\mathrm{rms,C}$, have about the same correlations. These are
strong with the block of basic NMP $E/A$ and $\rho_\mathrm{nm}$ plus
the isoscalar surface energy $a_\mathrm{surf}$. This confirms the
findings from figure \ref{fig:collect-NMP} where we see that radius
information has a large impact on the basic bulk binding. There are
some correlations with the binding energy of the superheavy element
Z=120/N=182 which is not a surprise as this nucleus also correlates with
the block $E/A$ and $\rho_\mathrm{nm}$. All other correlations are not
significant.

\subsection{Beyond linear analysis}
\label{sec:beyondlinear}

Standard $\chi^2$-analysis assumes that an observable $A(\mathbf{p})$
depends linearly on the model parameters $\mathbf{p}$ within the range
of reasonable $\mathbf{p}$.  We have checked that assumption by
carrying the expansion (\ref{eq:linear}) {for the observables} up
to quadratic terms {still assuming a
  quadratic form for $\chi^2$ as function of the parameters}. The
%\marginpar{{O}}
simple rules of integrating polynomials with Gaussians allow to
compute uncertainties and correlations also for this non-liner case in
straightforward manner.
\begin{figure}
\centerline{\epsfig{figure=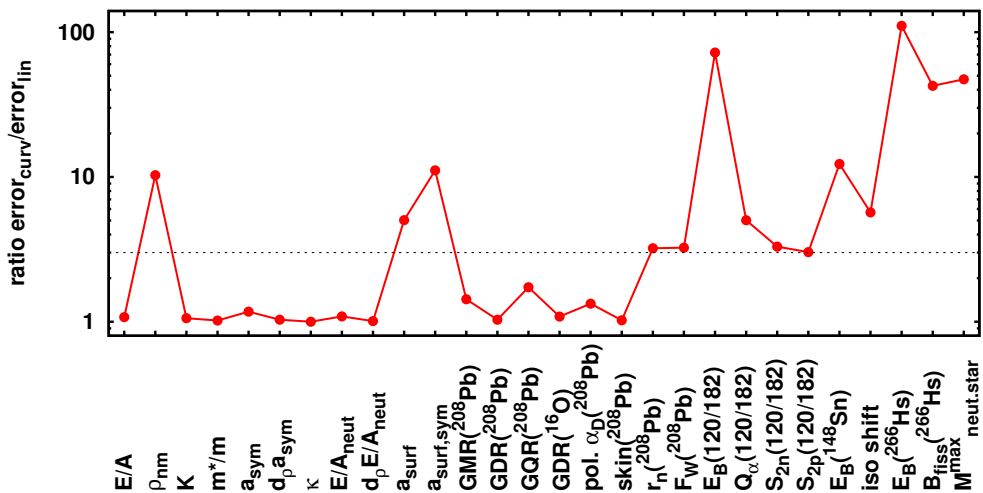,width=0.85\linewidth}}
\caption{\label{fig:collect-curv} Ratio of 
extrapolation uncertainties with curvature correction and without 
computed with SV-min. The horizontal line indicates a critical ratio of 
3 above which curvature effects become important.}
\end{figure}
Figure \ref{fig:collect-curv} shows the ratio of extrapolation
uncertainties computed from the quadratic expansion to those from the
linear model for a broad selection of observables.  We had also
checked the weight of the quadratic terms explicitly and this
delivers the same picture. Thus we take this ratio as a simple measure
of (non-)linearity.  The faint black horizontal line indicates the
limit up to which the assumption of linearity is acceptable. Most
observables are thus in the linear regime, a few of them reach into
the non-linear regime, and some of them (superheavy elements and
neutron stars) are dramatically non-linear with ratios going up to
100. However, this example has to be taken as an order of magnitude
estimate.  Such a highly non-linear requires a more careful evaluation
of correlation according to Spearman's analysis \cite{Sch07a}
which goes beyond the scope of this paper.

We have also checked the effect of non-linearity on covariances
$c_{AB}$ for SV-min.  The basic sorting into strongly correlated
combinations and weakly correlated ones is maintained.  It is only at
closer inspection that one can spot some changes of correlations if a
non-linear observable is involved. In most cases, non-linearity's
reduce correlations slightly.

\section{Conclusions and outlook}

We have explored the uncertainties inherent in the Skyrme-Hartree-Fock
(SHF) approach by employing, out of many, five different strategies
for error estimates: extrapolation error from $\chi^2$ analysis,
covariances between pairs of observables from $\chi^2$ analysis,
trends with systematically varied model parameters (practically
nuclear matter parameters), trends of residual errors, and block-wise
variation of fit data. For the latter strategy, we have fitted a
couple of new parametrizations where differing groups of fit
observables had been omitted, once the r.m.s. radii, once the surface
thickness, once both formfactor observables (surface, diffraction
radius), and finally all form information (surface, diffraction
radius, r.m.s. radius). This makes, together with the full fit, five
parametrizations which are then used in combination with all further
analysis. Some of the strategies yield similar information (e.g. trend
with parameters and covariances), however from different perspectives
which makes it useful to consider both.  In any case, the combination
of strategies is more informative than any single strategy alone.  Out
of the many interesting aspects worked out in the above studies, we
emphasize here a few prominent findings and indicate the directions
for further development:
\begin{enumerate}
\item\label{it:bind}
Fits only to binding energy (omitting any radius information) yield
already a very acceptable description of nuclear properties.  The
error on diffraction radius and surface thickness grows by factor
3--4, but remains with about 0.08 fm in an acceptable range. However,
the uncertainties in extrapolations can grow large
which proves the usefulness of having radius information in the fit.
Radius information is strongly correlated with bulk binding
(equilibrium energy and density) as well as surface energy
and it helps to fix theses quantities.
\item
We have found a conflict between the description of r.m.s. radius and
diffraction radius. Fitting only one of the both spoils the other
one. Fitting both yields a compromise. The precision which can be
achieved is small (0.02--0.04 fm), but limited to that in the present
model. Further development work on SHF and the computation of radii is
required to harmonize the data.
\item
Extrapolations to exotic nuclei show, of course, increasing
uncertainties with increasing distance to the set of fit nuclei.  The
growth of errors is large for energies of $r$-process nuclei (factor
6--10) and moderate for energies of super-heavy elements (factor
2--3).  Errors on radii, on the other hand, remain even nearly
constant. This is related to the tight connection of radii to bulk
binding (see point \ref{it:bind} above).  
\item
The more dramatic extrapolation to neutron stars is plagued by much
larger uncertainty. Neutron matter is highly correlated to isovector
forces which are less well fixed by fits to existing nuclei.  The
variation of fit data shows that there are probably large systematic
errors beyond the statistical uncertainties.  In spite of the
generally large uncertainties, there is a ``magic'' region around
density 0.13 fm$^{-3}$ 
where all parametrizations yield surprisingly small uncertainties and very 
similar predictions. This effect deserves further investigation.
\item
We have checked the assumption of linear parameter dependence,
employed in standard statistical analysis, for many observables. Most
of them show sufficient linearity, but some deviate dramatically.
These are typically the observables in far extrapolations, exotic
nuclei and neutron stars. The impact of non-linearity on extrapolation
uncertainties and covariances has yet to be investigated.
\end{enumerate}

\bigskip
\noindent
Acknowledgments: This work was supported by the Bundesministerium
f\"ur Bildung und Forschung (BMBF) under contract number 05P09RFFTB.
\bigskip
\section*{References}
\bibliographystyle{iopart-num}
%{unsrt}
\bibliography{chi2analysis}

\end{document}